# Representation of fields in inhomogeneous structured waveguides

M.I. Ayzatsky

National Science Center Kharkov Institute of Physics and Technology (NSC KIPT), 61108, Kharkov, Ukraine

E-mail: mykola.aizatsky@gmail.com

In this work we present the results of a study of the possibility of using a homogeneous basis and a new generalization of coupled modes theory to describe non-periodic structured waveguides. It was shown that for the studied inhomogeneities the single wave approximation gives good results. Within the framework of single wave approximation, the fields are represented as the sum of two components, one of which is associated with the forward eigen wave, and the second with the backward eigen wave. But this second component is not a backward wave. Because of the coupling, characteristics of this component are determined by the first component, which is significantly larger than the second.

## 1. INTRODUCTION

Waveguides that consist of similar (but not always identical) cells are called structured. Structured waveguides based on coupled resonators play an important role in many applications.

Recently it was proposed to use a uniform basis for description of non-periodic structured waveguides with ideal metal walls and a new generalization of the theory of coupled modes was constructed [1]. The obtained infinitive system of coupled equations transfers into the known system for periodic waveguides as the inhomogeneity tends to zero [2]. Before using proposed system of coupled equations to calculate waveguide characteristics, it is necessary to establish the possibility of reducing an infinitive system of equations. The best way to do this is to calculate the waveguide field distribution based on a rigorous electromagnetic code, find the expansion coefficients, and estimate how many terms need to be included to match the two field distributions and what is the accuracy of the matching.

This procedure requires solving two tasks: calculating the field distribution in non-periodic structured waveguides and calculating the eigen characteristics (propagation constants and eigen vectors) of periodic waveguides for various geometrical parameters. For the case of Disc-Loaded Waveguide (DLW) the Coupled Integral Equations Method (CIEM) (see, for example, [3,4]) is the most suitable, since the mentioned above problems can be solved on its bases [5]. The code CASCIE (Code for Accelerating Structures - Coupled Integral Equations) was used for calculating the field distribution in non-periodic structures [6]. We also developed a code that calculate the eigen characteristics of periodic DLWs. We used IMSL Fortran Numeric Library Routine EVCCG for computing the eigenvalues and eigenvectors of a complex matrices

## 2. FIELD EXPANSION

Electromagnetic fields in a non-periodic structured waveguide with ideal metal walls can be represented in the form of such series [1]

$$\vec{H}(\vec{r}) = \sum_{s>0} \left\{ C_s(z) \vec{H}_s^{(e,z)}(\vec{r}) + C_{-s}(z) \vec{H}_{-s}^{(e,z)}(\vec{r}) \right\}, \tag{1}$$

$$\vec{E}(\vec{r}) = \sum_{s>0} \left\{ C_s(z) \vec{E}_s^{(e,z)}(\vec{r}) + C_{-s}(z) \vec{E}_{-s}^{(e,z)}(\vec{r}) \right\} + \frac{\vec{j}_z}{i\omega\varepsilon_0\varepsilon} \tag{2}$$

where $\vec{E}_s^{(e,z)}(\vec{r}), \vec{H}_s^{(e,z)}(\vec{r})$ are modified eigen vector functions obtained by generalizing the eigen $\vec{E}_s^{(e)}, \vec{H}_s^{(e)}$ vectors of a homogeneous waveguide by special continuation of the geometric parameters (see Appendix 1 and [1]). The eigen waves of homogeneous waveguide we present as $(\vec{E}_{\pm s}, \vec{H}_{\pm s}) = (\vec{E}_{\pm s}^{(e)}, \vec{H}_{\pm s}^{(e)}) \exp(\gamma_{\pm s} z)$, where $(\vec{E}_{\pm s}^{(e)}, \vec{H}_{\pm s}^{(e)})$ are periodic functions of the z-coordinate. Expansions (1) and (2) differ slightly from those proposed in the previous work [1] by using the basis $\vec{E}_{\pm s}^{(e,z)}(\vec{r}), \vec{H}_{\pm s}^{(e,z)}(\vec{r})$ instead of $\vec{E}_{\pm s}^{(z)}(\vec{r}), \vec{H}_{\pm s}^{(z)}(\vec{r})$ (they differ in the factors $\exp(\gamma_{\pm s}^{(z)} z)$). Under such choice of the basis functions, the coefficients $C_s(z), C_{-s}(z)$ include an exponential dependence on the z-coordinate and, strictly speaking, cannot be named as amplitude. But we will use this term for the sake of simplicity of the further presentation.

Since $\vec{E}_{\pm s}^{(e,z)}(\vec{r}), \vec{H}_{\pm s}^{(e,z)}(\vec{r})$ are orthogonal [1], from (1) and (2) we obtain

$$N_s^{(z)} C_s(z) = \int_{S'} \left( \left[ \left( \vec{E} - \frac{\vec{j}_z}{i\omega\varepsilon_0\varepsilon} \right) \vec{H}_{-s}^{(e,z)} \right] - \left[ \vec{E}_{-s}^{(e,z)} \vec{H} \right] \right) \vec{e}_z dS,$$

$$N_s^{(z)} C_{-s}(z) = -\int_{S'} \left( \left[ \left( \vec{E} - \frac{\vec{j}_z}{i\omega\varepsilon_0\varepsilon} \right) \vec{H}_s^{(e,z)} \right] - \left[ \vec{E}_s^{(e,z)} \vec{H} \right] \right) \vec{e}_z dS, \tag{3}$$

where

$$N_s^{(z)}(z) = \int_{S_t} \left( \left[ \vec{E}_s^{(e,z)} \vec{H}_{-s}^{(e,z)} \right] - \left[ \vec{E}_{-s}^{(e,z)} \vec{H}_s^{(e,z)} \right] \right) \vec{e}_z dS \tag{4}$$

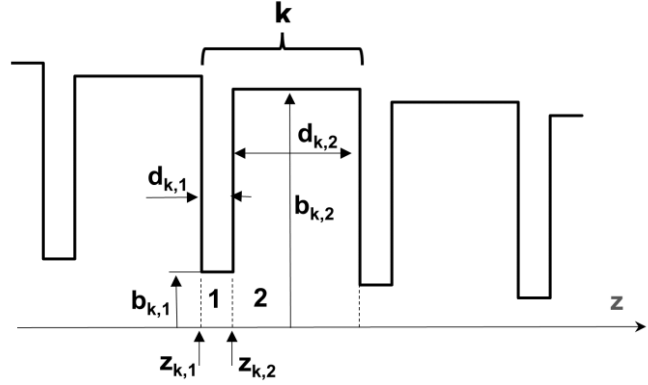

Figure 1

In this paper we will consider axisymmetric TH (E) electromagnetic fields in the finite structures on the base of inhomogeneous DLWs without current ($\vec{j}=0$). For a non-periodic circular DLW we will introduce the notion of the elementary cell with a number $k$ (see Figure 1), which starts at $z_{k,1} = \sum_{i}^{k-1}(d_{i,1}+d_{i,2})$ and consists of the first segment of a circular waveguide (region 1, $z_{k,1}<z<z_{k,1}+d_{k,1}=z_{k,2}$, length $d_{k,1}$, radius of waveguide $b_{k,1}$) and the second segment of a circular waveguide (region 2, $z_{k,2}<z<z_{k+1,1}$, radius $b_{k,2}$, length $d_{k,2}$).

In CIEM for the DLW with cylindrical geometry, electric and magnetic fields in the region $q$ ($q=1,2$) are represented in the form of series [5]

$$E_r^{(k,q)} = \sum_m \left\{ B_{m,1}^{(k,q)} \exp\left(\chi_m^{(k,q)}\tilde{z}\right) + B_{m,2}^{(k,q)} \exp\left(-\chi_m^{(k,q)}\tilde{z}\right) \right\} J_1\left(\frac{\lambda_m}{b_{k,q}}r\right), \tag{5}$$

$$E_z^{(k,q)} = -\sum_m \frac{\lambda_m}{\chi_m^{(k,q)} b_{k,q}} \left\{ B_{m,1}^{(k,q)} \exp\left(\chi_m^{(k,q)}\tilde{z}\right) - B_{m,2}^{(k,q)} \exp\left(-\chi_m^{(k,q)}\tilde{z}\right) \right\} J_0\left(\frac{\lambda_m}{b_{k,q}}r\right), \tag{6}$$

$$H_\varphi^{(k,q)} = i\frac{\omega b_{k,q}}{c}\frac{1}{Z_0} \sum_m \frac{\varepsilon}{\chi_m^{(k,q)} b_{k,q}} \left\{ B_{m,1}^{(k,q)} \exp\left(\chi_m^{(k,q)}\tilde{z}\right) - B_{m,2}^{(k,q)} \exp\left(-\chi_m^{(k,q)}\tilde{z}\right) \right\} J_1\left(\frac{\lambda_m}{b_{k,q}}r\right), \tag{7}$$

where $\left(\chi_m^{(k,q)}\right)^2 = \left(\frac{\lambda_m}{b_{k,q}}\right)^2 - \left(\frac{\omega}{c}\right)^2 \varepsilon$, $\tilde{z} = z - z_{k,q}$.

Coefficients $B_{m,1}^{(k,q)}, B_{m,2}^{(k,q)}$ are constant inside the cell and are found by equating the fields on the dividing surfaces. In addition to the standard division of the structured waveguide by interfaces between the adjacent regions, we use additional interfaces in places where electric field has the simplest transverse structure [5]. Moreover, the system of coupled integral equations is formulated for longitudinal electrical fields in contrast to the standard approach where the transverse electrical fields are unknowns. The final equations are a system of coupling matrix equations for expansion coefficients of the longitudinal electric field at these additional interfaces. Knowing expansion coefficients, we can calculate $B_{m,1}^{(k,q)}, B_{m,2}^{(k,q)}$ and, therefore, the field distribution.

Such approach gives us a simple procedure for finding eigen vectors of homogeneous waveguides. Instead of solving the system of coupling matrix equations, we must solve the eigen matrix problem, find eigen values and matrix eigen vectors and for selected eigen vector find coefficients $B_{m,1}^{(q,s)}, B_{m,2}^{(q,s)}$. The field distribution for the eigen wave we find using expressions (5)-(7) where $B_{m,1}^{(k,q)}, B_{m,2}^{(k,q)}$ are changed to $B_{m,1}^{(q,s)}, B_{m,2}^{(q,s)}$. After introducing a continuation of the geometric parameters (see [1]), we can find $C_s(z), C_{-s}(z)$. Since $J_1(\lambda_m r/b_{k,q})$ is the orthogonal set of functions, the integrals in (3) are converted into sums.

Using the same approach to calculate the distributions of fields and eigen waves makes it possible to reduce the influence of calculation errors on the results of assessing the accuracy of the decomposition (1)-(2).

For inhomogeneous structured waveguides, the use of the concepts of propagation and evanescence becomes not entirely clear. In proposed approach we use functions that are determined by a homogeneous waveguide and, therefore, $\vec{E}_s^{(e,z)}(\vec{r}), \vec{H}_s^{(e,z)}(\vec{r})$ can be classify as propagating or evanescent on the base of examining the parameter $\gamma_{\pm s}^{(z)}(z)$. But such classification cannot give the conformation that energy do not propagate through the considered cross-section. Only direct calculation of the power flow and detailed analysis can reveal the reflection or transport of energy, since when losses are taken into account, the power flow is nowhere equal to zero.

We will be interesting in possibility of using only two terms in the sums (1) and (2) – s=1, when we have a single-mode representation

$$\vec{E}_1(\vec{r}) = \vec{E}_1^+(\vec{r}) + \vec{E}_1^-(\vec{r}) = C_1(z)\vec{E}_1^{(e,z)}(\vec{r}) + C_{-1}(z)\vec{E}_{-1}^{(e,z)}(\vec{r}). \qquad (8)$$

Comparing results of calculation with using this approximation with the results obtained on the basis of coupling matrix equations [6], we can make the conclusion of possibility of using the single mode description (8).

We investigated the electromagnetic field expansion (1),(2) for structures which consist of $(K-2)$ cells and two coupler cells (the general number of cells equals $K$): one upstream (input) coupler and the other downstream (output) coupler. The couplers are connected to input and output infinitive cylindrical waveguides [6]. We also suppose that in the left semi-infinite waveguide the $TM_{0,1}$ eigen wave with unit amplitude propagates towards the considered section $E_z = J_0(\lambda_{01} r / b_w)\exp(i h_w z)$. Therefore, bellow all field values will correspond to this amplitude.

We will consider the frequency that lays inside (or close to) the main passband of $E_{01}$ wave in any cross-section of the waveguide. Everywhere below we will assume that the operating frequency is constant and equals $f = 2.856$ GHz. The lengths of regions 1 and 2 (see Figure 1) were also constant[1] and equal $d_{k,1} = d_1 = 0.5842$ cm and $d_{k,2} = d_2 = 2.9147$ cm, respectively, so the cell length $D = (d_1 + d_2)$ satisfies the condition $\omega D/c = 2\pi/3$. Energy absorption in the walls was modeled by introducing a complex dielectric into each cell ($\varepsilon = 1+i7.3E-5$) [5].

2.1 Homogeneous structures

The geometrical parameters of regular cells were chosen to satisfy the condition $\gamma_1 D = i2\pi / 3$. For solving the eigen matrix problem we used IMSL Fortran Numeric Library Routine EVCCG.

The couplers were tuned to homogeneous DLWs with irises radii $b_{k,1} = b_{I,1} = 1.381$ cm and $b_{k,1} = b_{II,1} = 1.02$ cm, respectively. The tuning procedure was based on the assumption that fields in a homogeneous DLW can be strictly described by the single mode approach (8) (see results below). The downstream coupler's geometrical sizes ($b_{K+1,1}, b_{K,2}$) were selected from the condition $|C_{-1}(z_c)/C_1(z_c)| < 10^{-3}$, where $z_c = D(K-4) + d_2/2$ (the middle of the $(K-4)$ cell).

It is known that in lossy DLWs the input coupler has slightly different geometric dimensions than the output coupler [7,8]. Therefore, only after tuning the output coupler the input coupler can be tuned by minimizing the reflectance of the entire structure. Two input couplers were tuned and such reflection coefficients were achieved: $R_{in} \approx$ 1.E-4 ($|C_{-1}/C_1| \approx$ 1.E-5, $b_{I,1} = 1.381$ cm) and $R_{in} \approx$ 1E-4 ($|C_{-1}/C_1| \approx$ 1.E-4, $b_{II,1} = 1.02$ cm).

Results of coupler tuning and accuracy of the single mode description (8) for the structure with 40 cells ($K = 40$, $b_{k,1} = b_{I,1} = 1.02$ cm $k = 2,...,39$) are presented in Figure 2-Figure 7. It can be seen that inside the regular part of section (cells $k = 4,...,37$) the accuracy of the single mode description is quite good.

Relative error of the representation of the longitudinal electric field $E_z(r = 0, z)$, calculated with using the CASCIE code, by the field of the single mode approach is less than 1.E-8 (see Figure 4). Relative error in the representation of $E_z(r = 0, z)$ by the field $E_{z,1}^+(r = 0, z)$ is determined by the reflection from the output coupler (see Figure 5).

In the coupler cells the single mode description gives the error up to 10%. This is due to the fact that a single mode description does not take into account the evanescent higher modes. Moreover, the functions $C_1(z)$ and $C_{-1}(z)$ inside the coupler cells change very sharply which is determined by the fast variation of generalized eigen vectors $\vec{E}_{\pm 1}^{(e,z)}(\vec{r})$ in the coupler cells.

---

[1] In the proposed approach, it is possible to consider cases with non-constant $d_{k,1}$ and $d_{k,2}$. This case will be considered elsewhere.

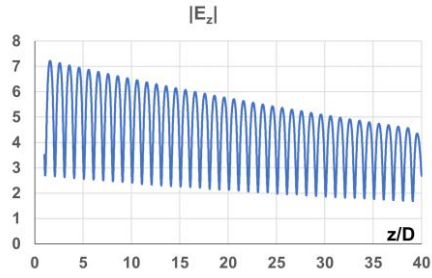
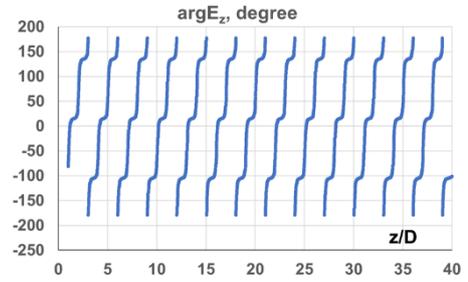

*Figure 2 Spatial distributions of the modulus of the longitudinal electric field $E_z(r=0,z)$ for a structure based on a homogeneous DLW with $b_{k,1}$ =1.02 cm, (CASCIE code)*

*Figure 3 Spatial distributions of the phase of the longitudinal electric field $E_z(r=0,z)$ for a structure based on a homogeneous DLW with $b_{k,1}$ =1.02 cm, (CASCIE code)*

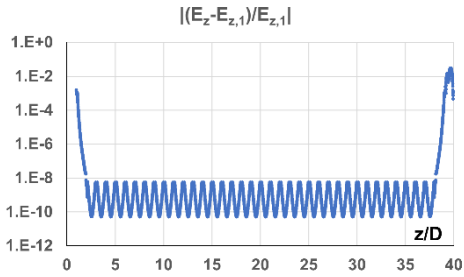
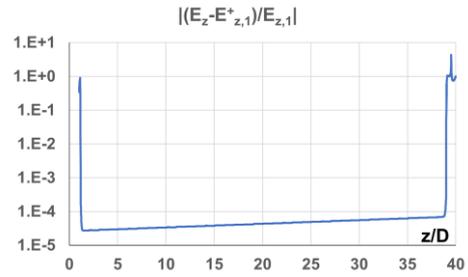

*Figure 4 Relative error in the representation of the longitudinal electric field $E_z(r=0,z)$, calculated on the base of the CASCIE code, by the field of the single mode approach $E_{z,1}(r=0,z)$ for a structure based on a homogeneous DLW with $b_{k,1}$ =1.02 cm*

*Figure 5 Relative error in the representation of the longitudinal electric field $E_z(r=0,z)$, calculated on the base of the CASCIE code, by the field $E_{z,1}^+(r=0,z)$ for a structure based on a homogeneous DLW with $b_{k,1}$ =1.02 cm*

Results of calculations, presented in Figure 6 and Figure 7, confirm that the used procedure (3) provides the correct way to expand the field into forward and backward waves in the case of homogeneous waveguide.

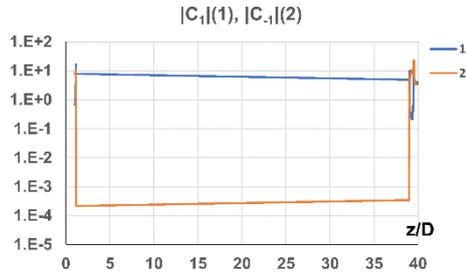
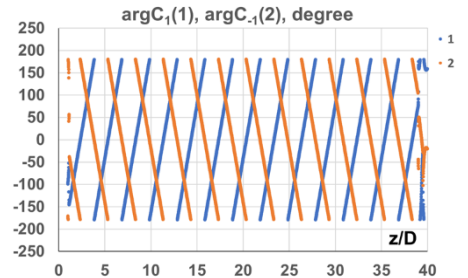

*Figure 6 Spatial distributions of the modulus of amplitudes $|C_1|$ (1) and $|C_{-1}|$ (2) for a structure based on a homogeneous DLW with $b_{k,1}$ =1.02 cm*

*Figure 7 Spatial distributions of the argument of amplitudes $\arg C_1$ (blue) and $\arg C_{-1}$ (red) for a structure based on a homogeneous DLW with $b_{k,1}$ =1.02 cm*

2.2 Inhomogeneous structures

Knowing the geometric dimensions of the couplers for two homogeneous DLW (type I and type II), we can study various inhomogeneous structures consisting of $K$ cell and having several homogeneous cells ($k=2\div K_S$) of type I at the beginning and several homogeneous cells ($k=K_E\div K-1$) of type II at the end of the section. For such construction there is no necessity to tune new couplers for different laws of inhomogeneity. In this case the values of disk apertures $b_{k,1}$ of the considered structures can be written as

$$b_{k,1} = \begin{cases} b_{I,1}, & k=2,...,K_S, \\ f(z), \ z=z_{k,1}=D\,k; & k=K_S+1,...,K_E, \\ b_{II,1}, & k=K_E+1,...,K-1, \end{cases} \qquad (9)$$

where the function $f(z)$ satisfied the following conditions

$$f(D(K_S+1))=b_{I,1},$$
$$f(DK_E)=b_{II,1}.$$
(10)

We have considered two cases:
1- $f(z)$ is a linear function;
2- $f(z)$ is a 5$^{th}$ degree polynomial with additional conditions $f'(D(K_S+1))=0$, $f'(DK_E)=0$ and $f''(D(K_S+1))=0$, $f''(DK_E)=0$.

The first distribution with $K \approx 40$ corresponds to accelerating sections with strong inhomogeneity, which were developed for acceleration high current with good efficiency (almost 100 % beam loading) [9,10]. With $K \approx 80$ it is similar to SLAC structure [11]). The second function was chosen to realize a smoother transition between homogeneous and inhomogeneous regions.

It was shown that for uniform DLW with $\varphi = 2\pi/3$ for $0.005 < \beta_g = v_g/c < 0.02$ we can use such dependency of the cavity radii $b$ on the irises radii $a$ ( $[a]=cm, [b]=cm$ ) [6]

$$b = 0.1576a^2 - 0.1476a + 4.0652.$$
(11)

For the structures under study, the radius of the cavity $b_{k,2}$ was selected using the following formulas

$$b_{k,2,0} = \begin{cases} b_{I,2}, & k=2,...,K_S, \\ 0.1576\left(\dfrac{b_{k,1}+b_{k+1,1}}{2}\right)^2 - 0.1476\left(\dfrac{b_{k,1}+b_{k+1,1}}{2}\right) + 4.0652, & k=K_S+1,...,K_E-1, \\ b_{II,2}, & k=K_E,...,K-1, \end{cases}$$
(12)

$$b_{k,2} = b_{k,2,0}(q_2 k + q_1), \quad k = K_S+1,...,K_E-1,$$
(13)

where $q_1$ and $q_2$ are found from the conditions $q_2 K_S + q_1 = b_{I,2}/b_{K_S,2,0}$ and $q_2 K_E + q_1 = b_{II,2}/b_{K_E,2,0}$ .

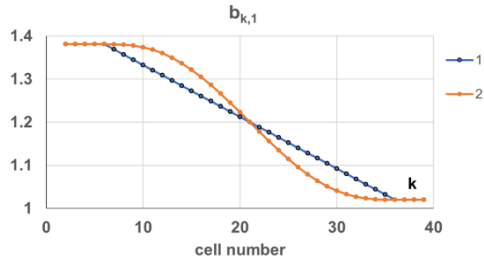
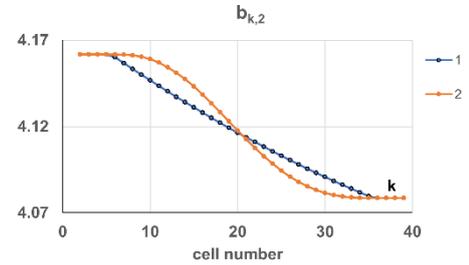

*Figure 8 Distributions of disk hole radii in structures with different inhomogeneities*

*Figure 9 Distributions of cell radii in structures with different inhomogeneities*

For sections with 40 cells ($K=40$, $K_S=5$ and $K_E=36$) the distributions of the radii of disk holes and the radii of cells with inhomogeneities of types 1 and 2 are presented in Figure 8 and Figure 9. The reflection coefficients for these structures are $|R_{input}| = 3.\text{E-3}$, and $|R_{input}| = 2.7\text{ E-3}$ respectively.

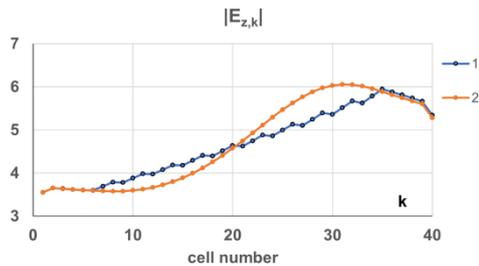
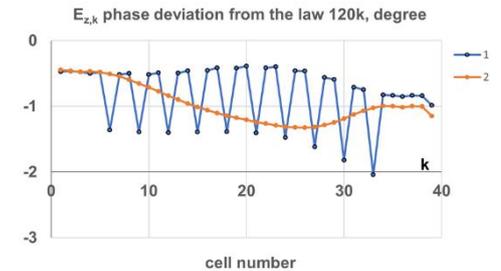

*Figure 10 Modulus of longitudinal electric field in the centers of cells ( $z = k(d_1+d_2)+d_2/2$, $r=0$ )*

*Figure 11 Deviation of phase distributions in the centers of cells from the law $120°k$*

Corresponding amplitudes and phases of longitudinal electric fields in the centers of cells, calculated using the code CASCIE, are given in Figure 10 and Figure 11. For a linear transition (case 1), small reflections from the region of the junction of homogeneous and inhomogeneous regions are observed. Deviation of the phase distribution from $2\pi k/3$ law is small and amounts to several degrees.

The main feature of the new approach in the coupling waves theory is introducing new basis functions $\vec{E}_s^{(z)} = \vec{E}_s\left(\vec{r}_\perp, z, b_1^{(z)}(z), b_2^{(z)}(z)\right)$, $\vec{H}_s^{(z)} = \vec{H}_s\left(\vec{r}_\perp, z, b_1^{(z)}(z), b_2^{(z)}(z)\right)$ which can be obtained by changing the constant values of geometric parameters $b_1, b_2$ in the eigen vectors $\vec{E}_s = \vec{E}_s\left(\vec{r}_\perp, z, b_1, b_2\right)$, $\vec{H}_s = \vec{H}_s\left(\vec{r}_\perp, z, b_1, b_2\right)$ of periodic waveguide at the set of functions $b_1^{(z)}(z), b_2^{(z)}(z)$ which satisfy certain conditions (see Appendix 1). For the geometry under consideration, we need to connect two horizontal lines with a function that has as many zero derivatives as possible at both ends of the segment. If we want to have $N$ zero derivatives, we can use a polynomial of degree ($2N$+1). Calculations shown that a choose $N>2$ don't improve the calculation accuracy. So, we used functions like this

$$b_1^{(z)}(z) = \begin{cases} b_{1,k}, & z_{k,1} < z < z_{k,2}, \\ b_{1,k} + \left(b_{1,k+1} - b_{1,k}\right)\left\{10\dfrac{\tilde{z}^3}{d_{2,k}^3} - 15\dfrac{\tilde{z}^4}{d_{2,k}^4} + 6\dfrac{\tilde{z}^5}{d_{2,k}^5}\right\}, \tilde{z} = z - z_{k,2}, & z_{k,2} < z < z_{k+1,1}, \end{cases} \quad (14)$$

$$b_2^{(z)}(z) = \begin{cases} b_{2,k-1} + \left(b_{2,k} - b_{2,k-1}\right)\left\{10\dfrac{\tilde{z}^3}{d_{1,k}^3} - 15\dfrac{\tilde{z}^4}{d_{1,k}^4} + 6\dfrac{\tilde{z}^5}{d_{1,k}^5}\right\}, \tilde{z} = z - z_{k,1}, & z_{k,1} < z < z_{k,2}, \\ b_{2,k}, & z_{k,2} < z < z_{k+1,1}, \end{cases} \quad (15)$$

Consequently, in the aria of the diaphragm the radius of the "resonator" $b_2^{(z)}(z)$ changes and in the aria of the "resonator" the radius of opening $b_1^{(z)}(z)$ changes.

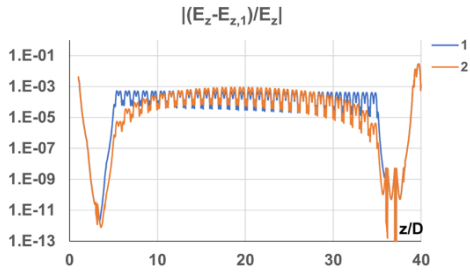

Figure 12 Relative error in the representation of the longitudinal electric field $E_z(r=0,z)$, calculated on the base of the CASCIE code, by the field of the single mode approach $E_{z,1}(r=0,z)$

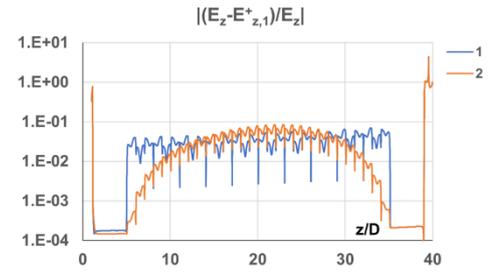

Figure 13 Relative error in the representation of the longitudinal electric field $E_z(r=0,z)$, calculated on the base of the CASCIE code, by the field $E_{z,1}^+(r=0,z)$

Results of calculations, presented in Figure 12 and Figure 14, show that under chosen parameters the single-mode approximation (8) gives a good accuracy of representation the field distribution in inhomogeneous structured waveguides. The relative error between the longitudinal electric field $E_z(r=0,z)$, calculated using the CASCIE code, and the field of the single mode approach $E_{z,1}(r=0,z)$ is less than 1E-3.

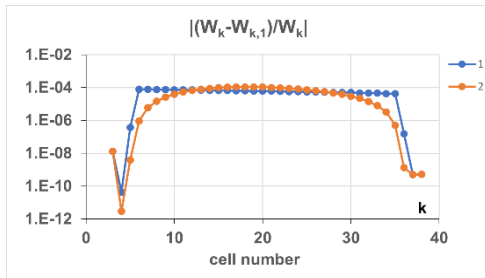

Figure 14 Relative error in the energy gain in a single cell calculated on the base of the longitudinal electric field $E_z(r=0,z)$ (CASCIE code) and using the field of the single mode approach $E_{z,1}(r=0,z)$

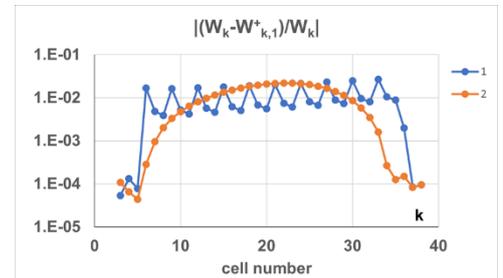

Figure 15 Relative error in the energy gain in a single cell calculated on the base of the longitudinal electric field $E_z(r=0,z)$ (CASCIE code) and using the field $E_{z,1}^+(r=0,z)$

For accelerator physics, an important characteristic is the amount of energy that a relativistic particle can receive when passing through the cell

$$W_k = e \operatorname{Re} \int_{z_{1,k}}^{z_{1,k+1}} E_z(z) \exp(-i\omega z/c) dz. \tag{16}$$

The relative error in calculating this parameter is even smaller - less than 1E-4 (see Figure 14). As follows from expression (8), the single mode field is divided into two components: the first, associated with the forward eigen wave and the second, associated with the backward eigen wave. If we limit ourselves to only the first part of the field (forward single mode approximation), then the error increases significantly and for inhomogeneities under consideration can reach several percent (see Figure 13 and Figure 15). For more steeper inhomogeneities, the errors will increase.

From these results it follows that the second component in the single-mode approximation can play an important role in the field representation in structured waveguides. This component is related with the backward eigen wave, but it doesn't represent the backward wave. Indeed, if the "amplitudes" $C_1$ of the first component behaves as an inhomogeneous wave travelling forward (see Figure 16 and Figure 17, phase grows linearly with the z coordinate) then the "amplitudes" of the second term $C_{-1}$ doesn't behave as an inhomogeneous wave travelling backward. Only in the homogeneous segments of waveguide (at the beginning and end of the waveguide) the dependence of phase $C_{-1}$ on $z$ coordinate has a decreasing character. $C_{-1}$ has a complex phase variation in the inhomogeneous segment: non-monotonic in the resonator and strong increasing across the disc (see Figure 19, Figure 21). This behavior is due to the coupling of two "amplitudes" (see Appendix 2) and will be explained elsewhere.

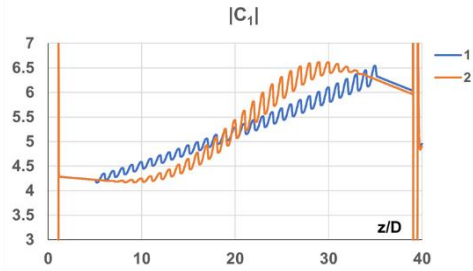

*Figure 16 Dependence of modulus $C_1$ on z-coordinate*

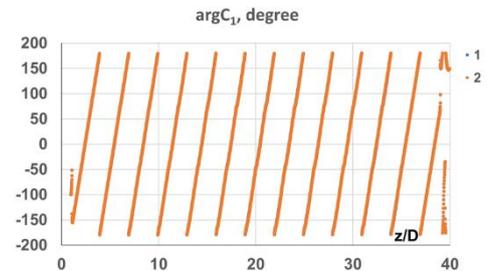

*Figure 17 Dependence of argument $C_1$ on z-coordinate*

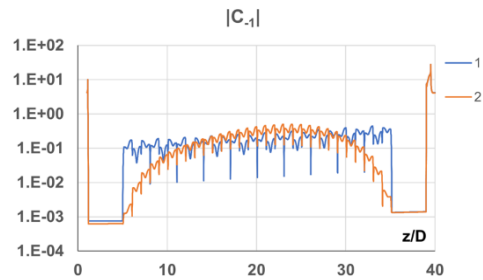

*Figure 18 Dependence of modulus $C_{-1}$ on z-coordinate*

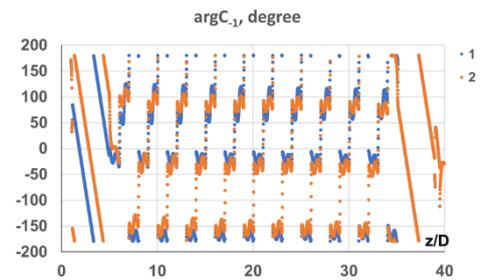

*Figure 19 Dependence of argument $C_{-1}$ on z-coordinate*

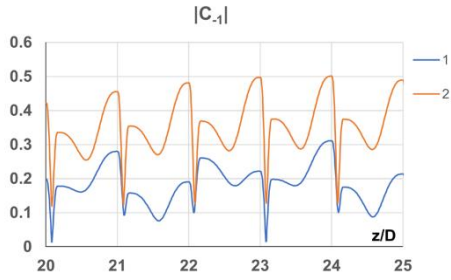

*Figure 20 Dependence of modulus $C_{-1}$ on z-coordinate*

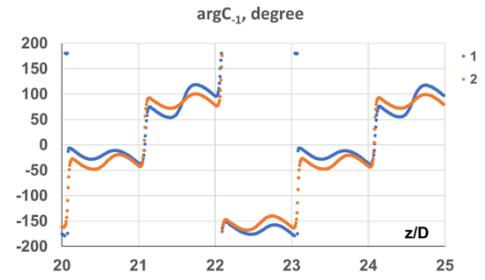

*Figure 21 Dependence of argument $C_{-1}$ on z-coordinate*

If the gradient of inhomogeneities decrease, then the representation errors will decrease too. It is confirmed by results of calculations that are presented in Figure 22-Figure 25 for the case of 86 cavities and the linear distribution of $b_{k,1}$. For this structure the gradient of inhomogeneities is two times smaller than in the case of 40 cavities structure. Relative error in the energy gain in a single cell calculated on the base of the longitudinal electric field $E_z(r=0,z)$ (CASCIE code) and using the field $E_{z,1}^+(r=0,z)$ is also reduced by almost half (compare Figure 15, Figure 13 and Figure 25).

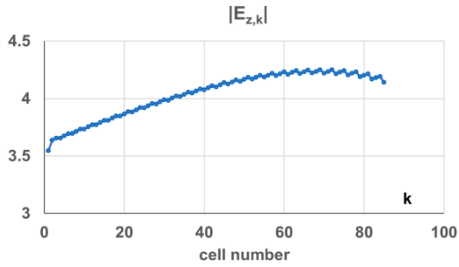

*Figure 22 Modulus of longitudinal electric field in the centers of cells ( $z = k(d_1+d_2)+d_2/2$, $r=0$ )*

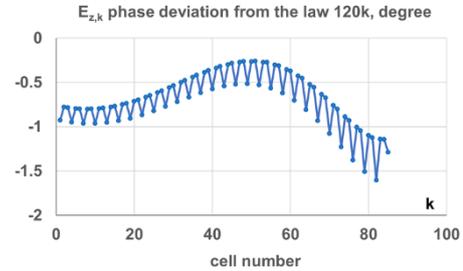

*Figure 23 Deviation of phase distributions in the centers of cells from the law $120° k$*

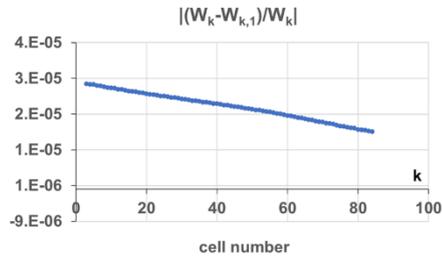

*Figure 24 Relative error in the energy gain in a single cell calculated on the base of the longitudinal electric field $E_z(r=0,z)$ ( CASCIE code) and using the field of the single mode approach $E_{z,1}(r=0,z)$*

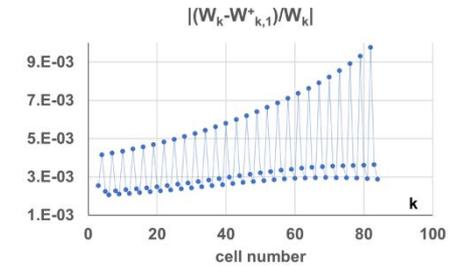

*Figure 25 Relative error in the energy gain in a single cell calculated on the base of the longitudinal electric field $E_z(r=0,z)$ ( CASCIE code) and using the field $E_{z,1}^+(r=0,z)$*

Presented above results confirm that the modified homogeneous eigen waves can be used as a basis for representation of electromagnetic fields in inhomogeneous structured waveguides. Using this basis, we can get a system of coupling equations which correctly describes non-periodic structured waveguides (Appendix 2). For some inhomogeneities we can reduce this system and use the single mode approximation.

## 2. CONCLUSIONS

New generalization of the theory of coupled modes, proposed in [1], gives possibility to describe non-periodic structured waveguides. Based on a set of eigen waves of a homogeneous periodic waveguide, a new basis of vector functions is introduced that takes into account the non-periodicity of the waveguide. Representing the total field as the sum of these functions with unknown scalar coefficients, a system of coupled equations that determines the dependence of these coefficients on the longitudinal coordinate can be obtained. For some inhomogeneities we can reduce this system and use the single mode approximation.

In this work we showed that for the studied inhomogeneities the single wave approximation gives good results. Within the framework of single wave approximation, the fields are represented as sum of two components, one of which is associated with the forward eigen wave, and the second with the backward eigen wave. But this second component is not a backward wave. The distributed coupling of waves due to the presence of distributed reflections leads to the fact that the characteristics of this component are determined by the characteristics of the first component, which is much larger than the second.

The calculation results showed that for the studied inhomogeneities the single wave approximation with taking into account the two components gives a good accuracy of field representation. The possibility of using only forward single mode approximation is determined by the magnitude of the gradient and the required representation accuracy.

## APPENDIX 1 EIGEN WAVES

We will assume that all quantities have a time variation given by $\exp(-i\omega t)$. Electromagnetic fields in a periodic structured waveguide are the solutions of the Maxwell's equations

$$rot\, \vec{E} = i\omega\mu_0 \vec{H}, \tag{17}$$

$$rot\, \vec{H} = -i\omega\varepsilon_0 \varepsilon \vec{E}, \tag{18}$$

with the periodic boundary conditions on the side metallic surface of the waveguide

This system has a set of solutions (eigen waves) $\left(\vec{E}_{\pm s}, \vec{H}_{\pm s}\right) = \left(\vec{E}_{\pm s}^{(e)}, \vec{H}_{\pm s}^{(e)}\right)\exp(\gamma_{\pm s} z)$, $s > 0$, $\gamma_{-s} = -\gamma_s$. Eigen vectors $\left(\vec{E}_{\pm s}^{(e)}, \vec{H}_{\pm s}^{(e)}\right)$ are the solutions of such equations

$$\begin{aligned} rot\, \vec{E}_{\pm s}^{(e)} + \gamma_{\pm s}\left[\vec{e}_z \vec{E}_{\pm s}^{(e)}\right] &= i\omega\mu_0 \vec{H}_{\pm s}^{(e)}, \\ rot\, \vec{H}_{\pm s}^{(e)} + \gamma_{\pm s}\left[\vec{e}_z \vec{H}_{\pm s}^{(e)}\right] &= -i\omega\varepsilon\varepsilon_0 \vec{E}_{\pm s}^{(e)} \end{aligned} \tag{19}$$

with the periodic boundary conditions and satisfy the orthogonality condition

$$\int_{S_\perp(z)} \left\{\left[\vec{E}_{\pm s}^{(e)} \vec{H}_{\mp s'}^{(e)}\right] - \left[\vec{E}_{\mp s'}^{(e)} \vec{H}_{\pm s}^{(e)}\right]\right\}\vec{e}_z dS = \begin{cases} 0, & s \neq s', \\ \pm N_s, & s = s', \end{cases} \tag{20}$$

where $S_\perp(z)$ - any cross-section of the waveguide. $S_\perp(z)$ and $\left(\vec{E}_{\pm s}^{(e)}, \vec{H}_{\pm s}^{(e)}\right)$ are the periodic functions of the $z$ coordinate.

Suppose that the boundary of the periodic waveguide is fully determined by the finite set of the geometrical parameters $g_i$, $i = 1,...,I$. Eigen solutions $\left(\vec{E}_{\pm s}^{(e)}, \vec{H}_{\pm s}^{(e)}\right)$ and propagation constant $\gamma_{\pm s}$ depend on the geometry of boundary, therefore we can write $\vec{E}_{\pm s}^{(e)} = \vec{E}_{\pm s}^{(e)}\left(\vec{r}_\perp, z, g_1, ..., g_I\right)$, $\vec{H}_{\pm s}^{(e)} = \vec{H}_{\pm s}^{(e)}\left(\vec{r}_\perp, z, g_1, ..., g_I\right)$, $\gamma_{\pm s}(g_1, ..., g_I)$.

For each fixed $z$ there are a set of $g_i^{(l)}$ (we will call them local geometrical parameters), that determine the geometry of the cross section $S_\perp^{(z)}(z, g_i^{(l)})$. The remaining parameters we will call global $g_i^{(g)}$. The division of the geometrical parameters $g_i$ into local and global ones depends on $z$.

We can consider new vector functions $\vec{E}_{\pm s}^{(e,z)} = \vec{E}_{\pm s}^{(e)}\left(\vec{r}_\perp, z, g_1^{(z)}(z), ..., g_I^{(z)}(z)\right)$, $\vec{H}_{\pm s}^{(e,z)} = \vec{H}_{\pm s}^{(e)}\left(\vec{r}_\perp, z, g_1^{(z)}(z), ..., g_I^{(z)}(z)\right)$, where $g_i^{(z)}(z)$ and its derivatives are continuous functions of $z$. We also introduce new function $\gamma_s^{(z)} = \gamma_s\left(g_1^{(z)}(z), ..., g_I^{(z)}(z)\right)$ The set $g_i^{(z)}(z)$ doesn't describe any real waveguide. But for each fixed $z$ the vectors $\vec{E}_{\pm s}^{(e,z)}$, $\vec{H}_{\pm s}^{(e,z)}$ represent the fields of a periodic waveguide (the virtue waveguide [12]) in the cross section $S_\perp^{(z)}(z, g_i^{(l,z)}(z))$, where $g_i^{(l,z)}(z)$ are the set of local geometrical parameters.

The vector functions $\vec{E}_{\pm s}^{(e,z)}$, $\vec{H}_{\pm s}^{(e,z)}$ are no longer the solutions to Maxwell equations. Indeed, as

$$\frac{\partial \vec{E}_s^{(e,z)}}{\partial z} = \frac{\partial \vec{E}_{s,g}^{(e,z)}}{\partial z} + \sum_i \frac{\partial \vec{E}_s^{(e,z)}}{\partial g_i} \frac{dg_i}{dz}, \tag{21}$$

where $\vec{E}_{s,g}^{(e,z)} = \vec{E}_s^{(e,z)}\left(\vec{r}, g_i = const\right)$, then

$$rot\, \vec{E}_s^{(e,z)} = i\omega\mu_0 \vec{H}_s^{(e,z)} - \gamma_s^{(z)}\left[\vec{e}_z \vec{E}_s^{(e,z)}\right] + \vec{E}_s^{(\nabla)}, \tag{22}$$

$$rot\, \vec{H}_s^{(e,z)} = -i\omega\varepsilon\varepsilon_0 \vec{E}_s^{(e,z)} - \gamma_s^{(z)}\left[\vec{e}_z \vec{H}_s^{(e,z)}\right] + \vec{H}_s^{(\nabla)}, \tag{23}$$

where

$$\vec{E}_s^{(\nabla)} = \sum_i \left[ \vec{e}_z \frac{\partial \vec{E}_s^{(e,z)}}{\partial g_i} \right] \frac{dg_i}{dz},$$
$$\vec{H}_s^{(\nabla)} = \sum_i \left[ \vec{e}_z \frac{\partial \vec{H}_s^{(e,z)}}{\partial g_i} \right] \frac{dg_i}{dz},$$
(24)

New vector functions $\vec{E}_s^{(e,z)}, \vec{H}_s^{(e,z)}$ still satisfy the boundary conditions on the contour $\vec{r}_\perp = \vec{r}_\perp^{(z)}(z, g_i^{(l,z)}(z))$, which limits the cross section $S_\perp^{(z)}(z, g_i^{(l,z)}(z))$ of some waveguide, and the orthogonality conditions

$$N_{\pm s, \mp s'}^{(z)} = \int_{S_t(z)} \left\{ \left[ \vec{E}_{\pm s}^{(e,z)} \vec{H}_{\mp s'}^{(e,z)} \right] - \left[ \vec{E}_{\mp s'}^{(e,z)} \vec{H}_{\pm s}^{(e,z)} \right] \right\} \vec{e}_z dS = \begin{cases} 0, & s \neq s', \\ \pm N_s^{(z)}\left(g_i^{(z)}(z)\right), & s = s', \end{cases}$$

## APPENDIX 2

The behavior of electromagnetic field is governed by Maxwell's equations

$$rot\, \vec{E} = i\omega \mu_0 \vec{H}, \tag{25}$$
$$rot\, \vec{H} = -i\omega \varepsilon_0 \varepsilon \vec{E} + \vec{j}. \tag{26}$$

We will look for a solution of equations (25) and (26) in the form of such a series

$$\vec{H} = \sum_{s>0} \left( C_s \vec{H}_s^{(e,z)} + C_{-s} \vec{H}_{-s}^{(e,z)} \right) \tag{27}$$

Using the vector relation

$$rot\left(C_s \vec{H}_s^{(e,z)}\right) = \frac{dC_s}{dz}\left[\vec{e}_z \vec{H}_s^{(e,z)}\right] + C_s \left(-i\omega\varepsilon_0 \vec{E}_s^{(e,z)} - \gamma_s^{(z)}\left[\vec{e}_z \vec{H}_s^{(e,z)}\right] + \vec{H}_s^{(\nabla)}\right) \tag{28}$$

and assuming that the series (27) can be differentiated term by term, we get from (26)

$$\vec{E} = -\frac{1}{i\omega\varepsilon_0\varepsilon} rot\, \vec{H} + \frac{\vec{j}}{i\omega\varepsilon_0\varepsilon} = \sum_{s>0}\left(C_s \vec{E}_s^{(e,z)} + C_{-s}\vec{E}_{-s}^{(e,z)}\right) + \frac{\vec{j}}{i\omega\varepsilon_0\varepsilon} -$$
$$-\frac{1}{i\omega\varepsilon_0\varepsilon}\sum_{s>0}\left\{\left(\frac{dC_s}{dz} - \gamma_s^{(z)}C_s\right)\left[\vec{e}_z \vec{H}_s^{(e,z)}\right] + \left(\frac{dC_{-s}}{dz} - \gamma_{-s}^{(z)}C_{-s}\right)\left[\vec{e}_z \vec{H}_{-s}^{(e,z)}\right] + \vec{H}_s^{(\nabla)}C_s + \vec{H}_{-s}^{(\nabla)}C_{-s}\right\} \tag{29}$$

Suppose that

$$\sum_{s>0}\left\{\left(\frac{dC_s}{dz} - \gamma_s^{(z)}C_s\right)\left[\vec{e}_z \vec{H}_s^{(e,z)}\right] + \left(\frac{dC_{-s}}{dz} - \gamma_{-s}^{(z)}C_{-s}\right)\left[\vec{e}_z \vec{H}_{-s}^{(e,z)}\right] + \vec{H}_s^{(\nabla)}C_s + \vec{H}_{-s}^{(\nabla)}C_{-s}\right\} = \vec{j}_\perp, \tag{30}$$

where $\vec{j} = \vec{j}_l + \vec{j}_\perp$. Then (29) takes the form

$$\vec{E} = \sum_{s>0}\left(C_s \vec{E}_s^{(e,z)} + C_{-s}\vec{E}_{-s}^{(e,z)}\right) + \frac{\vec{j}_l}{i\omega\varepsilon_0\varepsilon} \tag{31}$$

We cannot use in (30) the full current $\vec{j}$ as the sums don't have the longitudinal components. Substitution of (31) and (27) into the equation (25) gives

$$\sum_{s>0}\left\{\left(\frac{dC_s}{dz} - \gamma_s^{(z)}C_s\right)\left[\vec{e}_z \vec{E}_s^{(e,z)}\right] + \left(\frac{dC_{-s}}{dz} - \gamma_{-s}^{(z)}C_{-s}\right)\left[\vec{e}_z \vec{E}_{-s}^{(e,z)}\right] + \vec{E}_s^{(\nabla)}C_s + \vec{E}_{-s}^{(\nabla)}C_{-s}\right\} = -\frac{1}{i\omega\varepsilon_0\varepsilon} rot\, \vec{j}_l \tag{32}$$

Making some transformations with Eq.(30) and (32) [1,2], we obtain

$$N_s^{(z)}\left(\frac{dC_s}{dz} - \gamma_s^{(z)}C_s\right) + \frac{1}{2}\frac{dN_s^{(z)}}{dz}C_s + \sum_{s'}\left(C_{s'}W_{s',-s}^{(z)} + C_{-s'}W_{-s',-s}^{(z)}\right) = \int_{S_\perp^{(z)}} \vec{j}\vec{E}_{-s}^{(e,z)}dS, \tag{33}$$

$$N_s^{(z)}\left(\frac{dC_{-s}}{dz} - \gamma_{-s}^{(z)}C_{-s}\right) + \frac{1}{2}\frac{dN_s^{(z)}}{dz}C_{-s} - \sum_{s'}\left(C_{s'}W_{s',s}^{(z)} + C_{-s'}W_{-s',s}^{(z)}\right) = -\int_{S_\perp^{(z)}} \vec{j}\vec{E}_s^{(e,z)}dS, \tag{34}$$

where

$$W_{k',k}^{(z)} = \frac{1}{2}\sum_i \frac{dg_i^{(z)}}{dz}\int_{S_\perp^{(z)}(z)}\left\{\left[\frac{\partial \vec{E}_{k'}^{(e,z)}}{\partial g_i^{(z)}} \vec{H}_k^{(e,z)}\right] + \left[\frac{\partial \vec{E}_k^{(e,z)}}{\partial g_i^{(z)}} \vec{H}_{k'}^{(e,z)}\right] - \left[\vec{E}_k^{(e,z)} \frac{\partial \vec{H}_{k'}^{(e,z)}}{\partial g_i^{(z)}}\right] - \left[\vec{E}_{k'}^{(e,z)} \frac{\partial \vec{H}_k^{(e,z)}}{\partial g_i^{(z)}}\right]\right\}\vec{e}_z dS. \tag{35}$$

# REFERENCES

1 M.I. Ayzatsky Coupled-mode theory for non-periodic structured waveguides https://doi.org/10.48550/arXiv.2309.06280

2 Weinstein LA. Electromagnetic waves. Moscow: Radio i svayz; 1988. 440 p. (In Russian).

3 S. Amari, J. Bornemann, and R. Vahldieck, "Accurate analysis of scattering from multiple waveguide discontinuities using the coupled integral equation technique," J. Electromag. Waves Applicat., 1996, V.10, pp. 1623–1644.

4 S. Amari; R.Vahldieck, J. Bornemann, P. Leuchtmann Spectrum of Corrugated and Periodically Loaded Waveguides from Classical Matrix Eigenvalues. IEEE Transactions on Microwave Theory and Techniques, 2000, V. 48, N. 3, pp.453-460.

5 M.I. Ayzatsky Modification of coupled integral equations method for calculation the accelerating structure characteristics PAST, 2022, N.3, pp.56-61,. https://doi.org/10.46813/2022-139-056; https://doi.org/10.48550/arXiv.2203.035182022

6 M.I. Ayzatsky Fast code CASCIE (Code for Accelerating Structures -Coupled Integral Equations). Test Results, https://doi.org/10.48550/arXiv.2209.11291, 2022

7 D.H. Whittum Introduction to Electrodynamics for Microwave Linear Accelerators. In: S.I.Kurokawa, M.Month, S.Turner (Eds) Frontiers of Accelerator Technology, World Scientific Publishing Co.Pte.Ltd., 1999

8 M.I. Ayzatsky A novel approach to the synthesis of the electromagnetic field distribution in a chain of coupled resonators. PAST, 2018, N.3, pp.29-37

9 M.I.Ayzatsky, E.Z.Biller. Development of Inhomogeneous Disk-Loaded Accelerating Waveguides and RF-coupling. Proceedings of Linac96, 1996, v.1, p.119-121

10 E. Jensen CTF3 drive beam accelerating structures. Proceedings of Linac2002, p.34-36

11 R.B Neal, General Editor, The Stanford Two-Mile Accelerator, New York, W.A. Benjamin, 1968

12 S. G.Johnson, P.Bienstman, M. A.Skorobogatiy, M.Ibanescu, E.Lidorikis, J. D. Joannopoulos, Adiabatic theorem and continuous coupled-mode theory for efficient taper transitions in photonic crystals. Physical Review E, 2002, 66, 066608